\documentclass[epj]{svjour}
\usepackage{graphics}
\usepackage{graphicx}
\usepackage{amsmath}
\usepackage{xcolor}
\usepackage[labelfont=footnotesize, textfont=footnotesize]{caption}
\newcommand{\pol}[1]{\mathaccent"017E{#1}}
\begin{document}
\title{Vector analyzing powers in the $d(\pol{p},pp)n$ and $d(\pol{p},{\rm{^{2}He}})n$ channels at 135~MeV }

\author{H.~Tavakoli-Zaniani\inst{1,2}\thanks{h.tavakoli.zaniani@rug.nl} \and 
M.~Eslami-Kalantari\inst{2}\thanks{meslami@yazd.ac.ir}\and 
H. R. Amir-Ahmadi\inst{1}\and
M.~T.~Bayat\inst{1}\and  
A.~Deltuva\inst{3}\and
J.~Golak\inst{4}\and 
N.~Kalantar-Nayestanaki\inst{1}\and
 St.~Kistryn\inst{5}\and 
 A.~Kozela\inst{6}\and 
 H. Mardanpour\inst{1}\and
 J.~G.~Messchendorp\inst{1}\thanks{j.g.messchendorp@rug.nl}\and 
 M.~Mohammadi-Dadkan\inst{1,7}\and
 A. Ramazani-Moghaddam-Arani\inst{8}\and 
 R.~Ramazani-Sharifabadi\inst{1,9}\and 
 R.~Skibi{\'{n}}ski\inst{4}\and 
 E.~Stephan\inst{10}\and
 H.~Wita{\l}a\inst{4}
}

%
%
\institute{KVI-CART, University of Groningen, Groningen, the Netherlands \and
 Department of Physics, School of Science, Yazd University, Yazd, Iran \and 
 Institute of Theoretical Physics and Astronomy, Vilnius University, Lithuania \and 
 M.Smoluchowski Institute of Physics, Jagiellonian University, Krak$\acute{o}$w, Poland \and
 Institute of Physics, Jagiellonian University, Krak$\acute{o}$w, Poland \and 
 Institute of Nuclear Physics, PAS, Krak$\acute{o}$w, Poland \and 
 Department of Physics, University of Sistan and Baluchestan, Zahedan, Iran \and 
 Department of Physics, Faculty of Science, University of Kashan, Kashan, Iran \and
 Department of Physics, University of Tehran, Tehran, Iran \and 
 Institute of Physics, University of Silesia, Chorz$\acute{o}$w, Poland}

%
\date{Received: date / Revised version: date}
%
\abstract{
Vector analyzing powers, $A_x$ and $A_y$, of the proton-deuteron break-up reaction have been measured by using a polarized-proton beam at 135~MeV impinging on a liquid-deuterium target. For the experiment, the Big Instrument for Nuclear-polarization Analysis (BINA) was used at KVI, Groningen, the Netherlands. The data are compared to the predictions of Faddeev calculations using state-of-the-art two- and three-nucleon potentials. Our data are reasonably well described by calculations for the kinematical configurations at which the three-nucleon force (3NF) effect is predicted to be small. However, striking discrepancies are observed at specific configurations, in particular in the cases of symmetric configurations, where the relative azimuthal angle between the two protons is small which corresponds to the $d(\pol{p},{\rm{^{2}He}})n$ channel. The vector analyzing powers of these configurations are compared to the proton deuteron elastic scattering to study the spin-isospin sensitivity of the 3NF models. The results are compared to the earlier results of the proton-deuteron break-up reaction at 190~MeV proton-beam energy~\cite{Mardanpour2010}. A disagreement is observed for both proton-beam energies between data and calculations which points 
to a deficiency in the treatment of spin-isospin part of the 3NF.
} 
\maketitle
\section{Introduction}
\label{intro}
A detailed description of nuclear forces is essential for understanding the properties of nuclei and the dynamics in few-nucleon scattering processes. A large part of the interaction between nucleons is described in meson-exchange theories or in the framework of Chiral Perturbation-Theory (ChPT)~\cite{kolck94,Epelbaum2000,Epelbaum2002}. Impressive developments are made in describing the two-nucleon force (2NF) within lattice QCD \cite{Machleidt2017}, albeit that they are still computationally very expensive and only applicable in a few cases~\cite{Kostas2015}.

The need for an additional 3NF became evident when comparing three-body scattering observables and binding energies of light nuclei with the state-of-the-art calculations~\cite{pieper01}. All 2NF models such as CD-Bonn~\cite{Machleidt2001}, Argonne-V18 (AV18)~\cite{Wiringa1995}, Reid93, Nijmegen I, Nijmegen II~\cite{Stoks1993} give an excellent description of proton-proton and proton-neutron scattering and the properties of the deuteron, such as its binding energy~\cite{Nasser2012}. For the simplest three-nucleon system, the triton, a precise solution of the three-nucleon Faddeev equations employing only 2NF clearly underestimates the experimental binding energy. Similar discrepancies were observed in the binding energy of even heavier nuclei ~\cite{Friar1993}. The need for an additional  3NF also became clear through studying nucleon-deuteron scattering observables~\cite{Witala2009}. A first sign came from the differential cross section of the proton-deuteron elastic interaction~\cite{Shimizu1982}, where the description of the minimum of the cross section could only be improved by adding a 3NF to the potential. Unfortunately, the inclusion of 3NF only partly remedies these deficiencies.

Several theoretical formalisms have been developed, such as a dynamic $\Delta$~\cite{Hajduk1983} and the Tucson-Melbourne~\cite{Coon1979} 3NF. These were embedded within the rigorous calculations using the Faddeev-type equations by Krak$\acute{o}$w and Hannover-Lisbon theory groups. 

Another source of information about the three-nucleon systems is the nucleon-deuteron break-up process.
In this paper, the specific region that corresponds to the reaction $d(\pol{p},pp)n$ at 135~MeV is studied where the two final-state protons scatter with a small relative energy and a small opening angle. A comparison is made with data of the $\pol{p}d$ elastic reaction that were measured by BINA at KVI~\cite{Ramazani2008}. The aim is to study the spin-isospin dependence of $A_x$ and $A_y$ vector-analyzing powers. Moreover, the results are compared to proton-deuteron break-up data taken at similar configurations at a slightly higher proton-beam energy of 190~MeV~\cite{Mardanpour2010}. In the latter case, a spin-isospin anomaly in $A_y$ was reported. In this paper, we address the energy dependence of this observation and we present in addition new results for $A_x$.

\section{Experimental setup}
\label{Exp}
The $\pol{p}+d$ break-up reaction was studied using a polarized beam of protons with an energy of 135~MeV impinging on a liquid deuterium target which was located at the center of BINA vacuum chamber. The polarized beam was prepared by POLIS (POLarized Ion Source) and accelerated by AGOR (Accelerateur Groningen ORsay) at KVI, the Netherlands. The momenta of the final-state protons were measured using the various detectors of BINA. The forward-scattered protons of the break-up reaction were detected by the forward wall placed outside the vacuum.

\begin{figure}
\centering 
\includegraphics[scale=0.35]{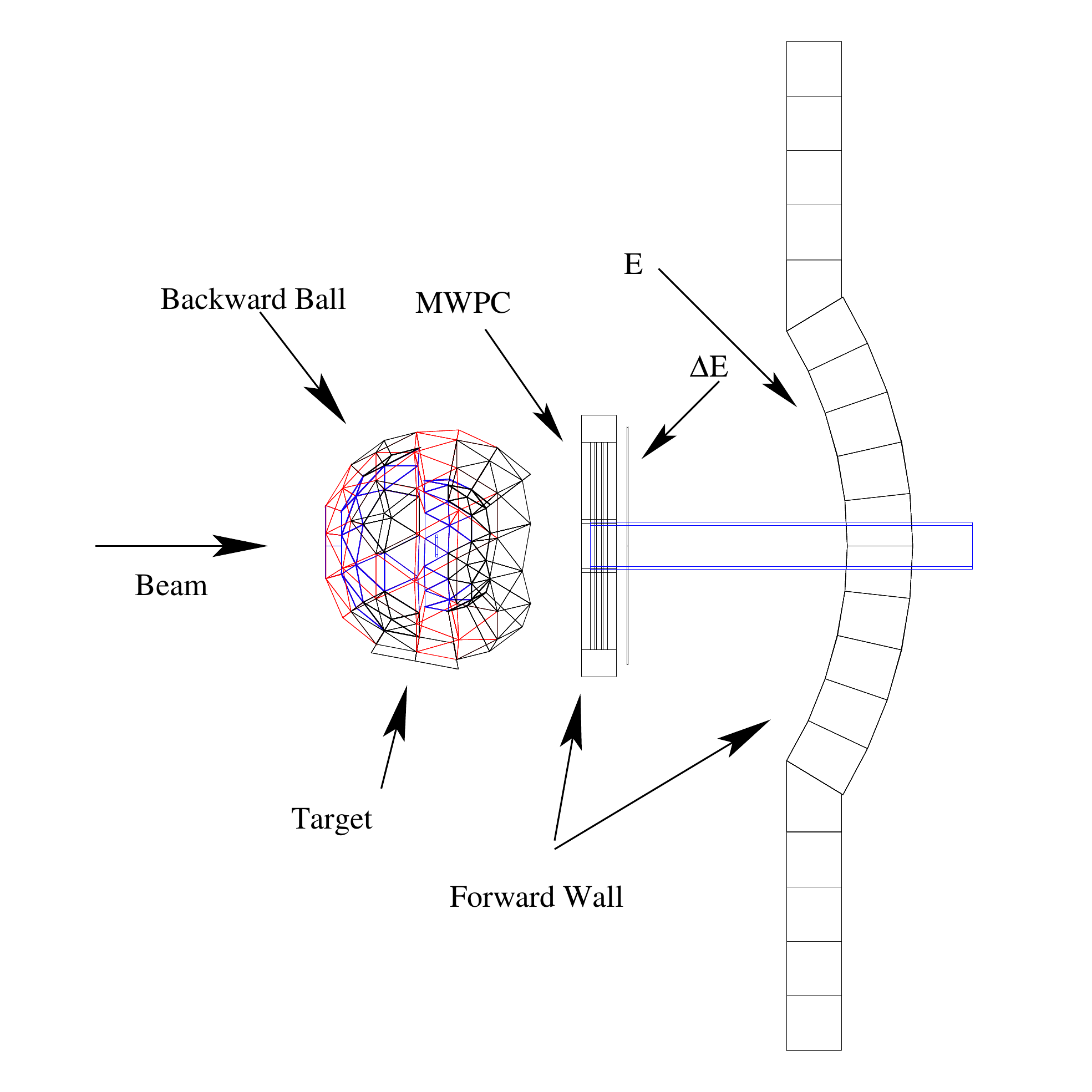} 
\caption{A schematic drawing of the forward-wall, the backward-ball, the beam direction and the target position of BINA.} 
\label{fig:bina} 
\end{figure}

As shown in Fig.~\ref{fig:bina}, the detector system BINA is composed of two main parts, the forward wall and the backward ball. The forward wall can measure the energy, the position, and the type of the particle at the polar scattering angles between $10^{\circ}-35^{\circ}$ and azimuthal angels between $0^{\circ}-360^{\circ}$. It has three parts, namely \textit{E}-scintillators,  $\Delta$\textit{E}-scintillators, and a Multi-Wire Proportional Chamber (MWPC). 

The forward-wall $E$-detector of BINA consists of a cylindrically-shaped array of plastic scintillators to measure the energy of the particles and two arrays of scintillators, up and down (wings), for detecting the secondary scattered particles for polarization-transfer experiments. For the present experiments, these wings were not used. The array of the $\Delta$\textit{E}-scintillators is composed of 12 parallel vertical plastic scintillator bars. The MWPC of BINA consists of 3 planes, X, Y, and U. The X plane is made of 236 parallel vertical wires, the Y plane has 236 horizontal wires, and the U plane has 296 parallel wires placed at an angle of $45^\circ$ with respect to the X or Y planes. Scattered particles with enough energy travel from the target to the scintillators. They pass through the MWPC and, as a result, their coordinates are recorded. Subsequently, particles pass through the $\Delta$\textit{E}-scintillators in which a small fraction of their energy is deposited. Finally, the particle is stopped (for protons with an energy less than 140~MeV) inside the \textit{E}-scintillators and its deposited energy is measured.

The backward part of BINA is covered by a ball-shaped detector.  The backward-ball is the scattering chamber and a detector at the same time. The incoming beam hits the target which is placed at the center of the ball. The particles are detected by the forward wall, which is placed outside the vacuum, and particles scattering to angles  larger than $35^{\circ}$ are detected by the ball. Tab.~\ref{table:tab1} shows the experimental features used in this experiment. 

\begin{table}
   \renewcommand\thetable{1}
   \centering
\caption{The experimental features of this experiment.}
\label{table:tab1}
\begin{tabular}{ll} 
 \hline
Beam particle & Polarized proton  \\
Energy &  135~MeV  \\
Average beam current &    15~pA  \\
Target &    Liquid deuterium  \\
Target thickness &   3.85~mm \\
Energy resolution & 5\% \\              
 \hline
\end{tabular}
\end{table}

\section{Data analysis}
\label{Analysis}
\begin{figure*}[tpb]
\centering
\includegraphics[scale=0.7]{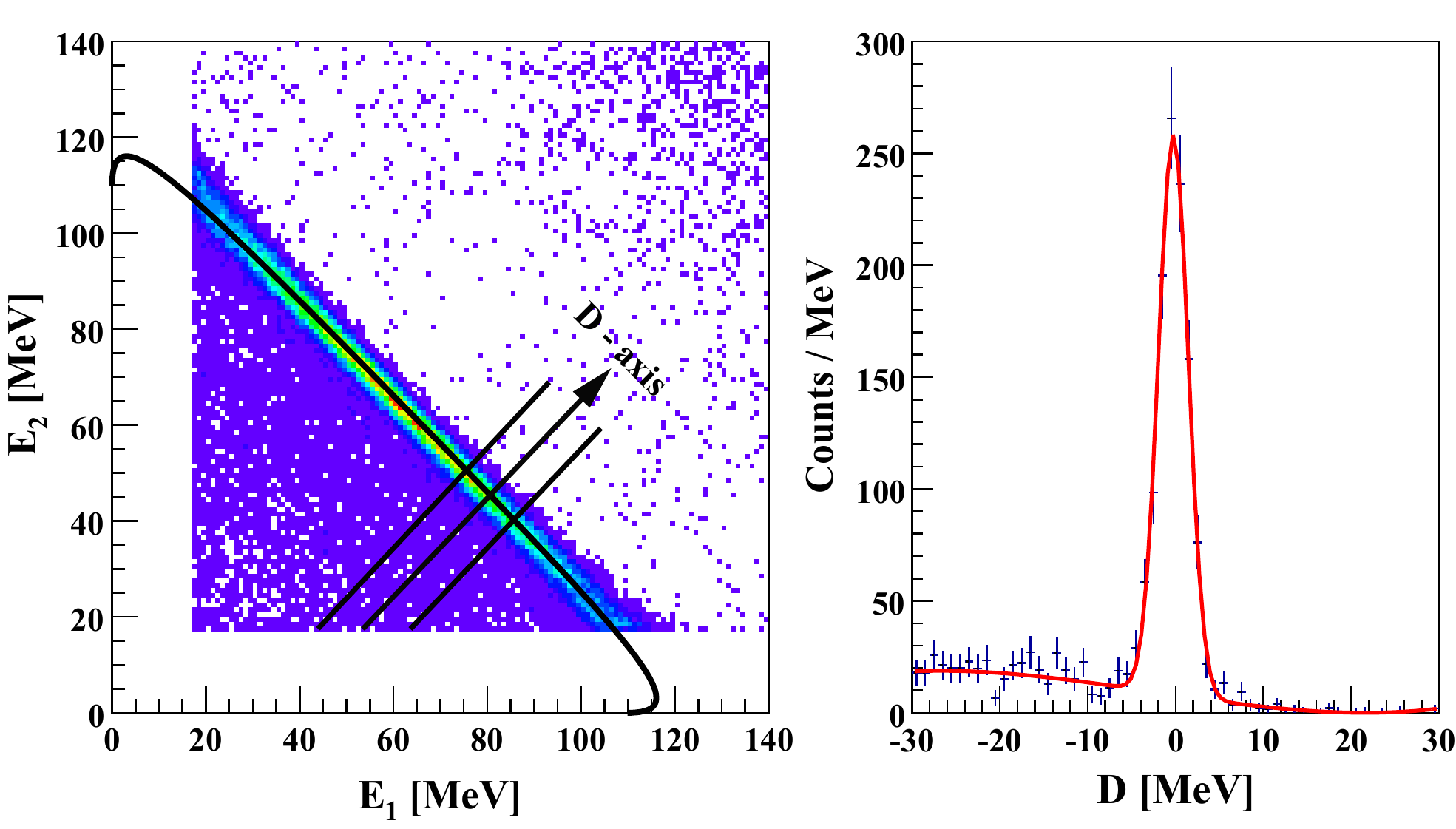}
\captionsetup{width=.85\linewidth, justification=justified}
\caption{The left panel shows $E_1$ versus $E_2$, spectrum of the two protons registered at $(\theta_{1}, \theta_{2}, \phi_{12})=(24^{\circ}\pm2^{\circ}, 24^{\circ}\pm2^{\circ}, 180^{\circ}\pm5^{\circ})$. The solid line shows the kinematical curve calculated for the central values of the experimental angular bins. The right panel is the projection of events along the D-axis for one slice shown on the left panel. The red line shows the result of a fit through the data. We refer to the text for more details. }
\label{fig:one}
\end{figure*}
Analyzing powers have successfully been measured as a function of the arclength, $S$, along the curve presenting the correlation between the energies of the final-state protons for different combinations of their polar and relative azimuthal angles $(\theta_{1}, \theta_{2}, \phi_{12}=\phi_{1}-\phi_{2})$.  The kinematic variable $S$ corresponds to the arclength along the kinematic curve with $S=0$ at the point where $E_{1}$ is at minimum. The left panel in Fig.~\ref{fig:one} shows the energy correlation between the two forward scattered protons after the energy calibration procedure. 
This procedure is performed by finding a calibration function that minimizes the shortest distance between the measured data points of the $\pol{p}d$ break-up reaction for all scintillators ($E_1$, $E_2$) to the kinematical $S$-curve. This is done via an iterative procedure. 
The data correspond to a specific kinematical configuration $(\theta_{1}, \theta_{2}, \phi_{12})=(24^{\circ}\pm2^{\circ}, 24^{\circ}\pm2^{\circ}, 180^{\circ}\pm5^{\circ})$. The solid line shows the kinematical $S$-curve calculated for the central values of the angular bins. The right panel in Fig.~\ref{fig:one} depicts the projection of events for one slice of $S-\Delta\textit{S}$ and $\textit{S}+\Delta\textit{S}$ with $\Delta\textit{S}=4.75$~MeV, along the D-axis taken perpendicular to the $S$-curve. The peak around zero corresponds to break-up events. Most of the events on the left-hand side of the peak are also due to break-up events, however in these cases, the protons have lost part of their energy due to hadronic interactions of the charged particles with the large material budget of the scintillators.
 The amount of accidental background is small as can been seen from the small amount of events on the right-hand side of the peak. We fit this spectrum by using a third-order polynomial, representing the hadronic interactions and accidental background, and a Gaussian function, representing the signal. The extracted number of signal events were corrected by the data-acquisition dead time, and the collected beam charge.

\begin{figure}[tpb] 
\includegraphics[scale=0.45]{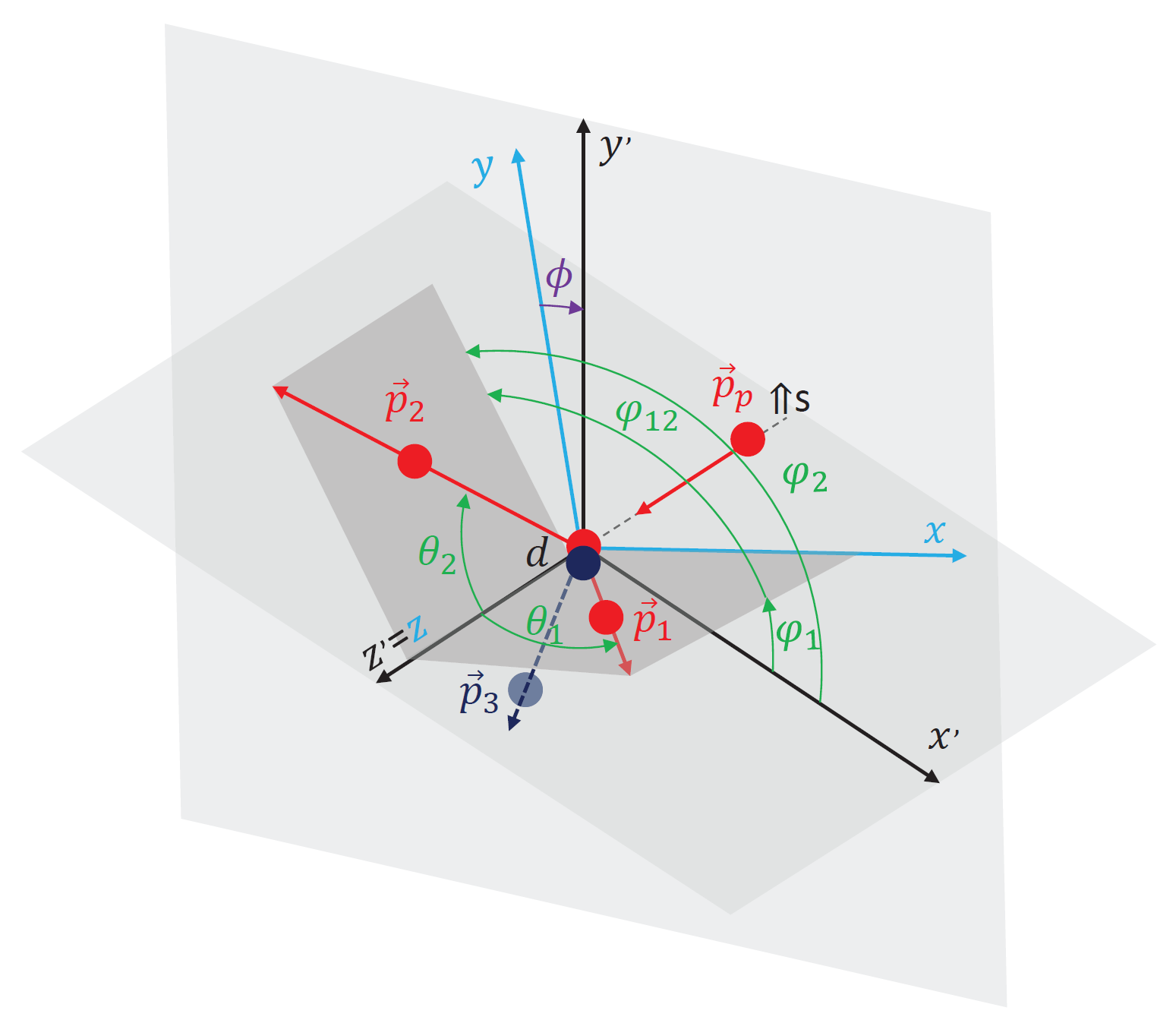}
\caption{Definition of the coordinate systems for
the break-up reaction with transversally-polarized beam: The laboratory system ($\textit{x}^\prime$, $y^\prime$, $z^\prime$), with $z^\prime$ along the beam-momentum direction and $y^\prime$ vertically upward, is used to define the angular configuration of the outgoing protons (with momenta $p_{1}$ and $p_{2}$). The reaction coordinate system is shown with $z$ along the beam-momentum direction and $x$ obtained by projection of $p_{1}$ onto the plane perpendicular to $z$. The angle $\phi$ is defined as the angle between the $y$ axis and the spin quantization axis $s$, which, in the studied case, is vertical (parallel to $y^\prime$)~\cite{Ohlsen1981,Stephan2010}.}
\label{fig:coor}
\end{figure}

\begin{figure*}
\centering
\includegraphics[scale=0.4]{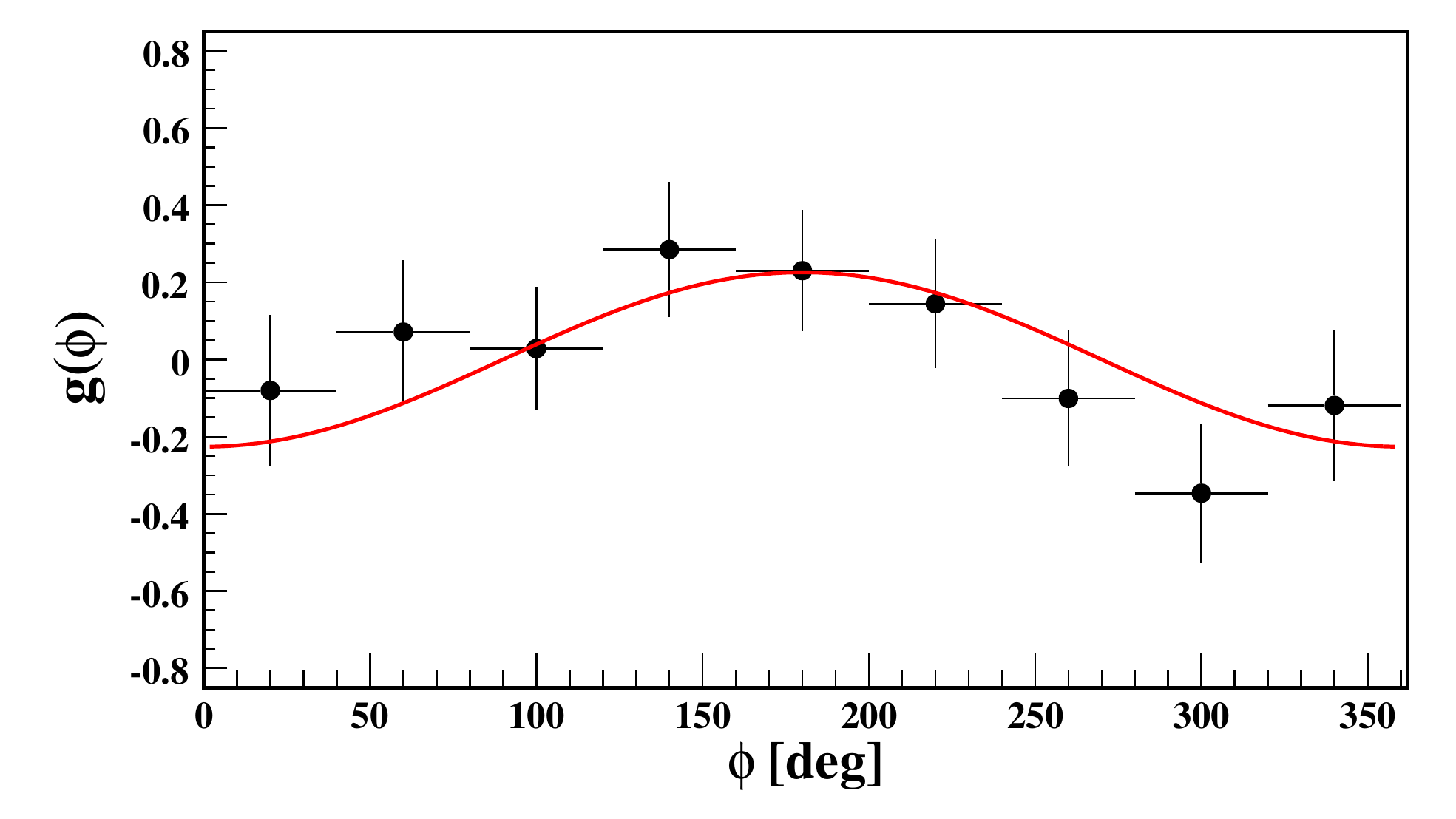} 
\includegraphics[scale=0.4]{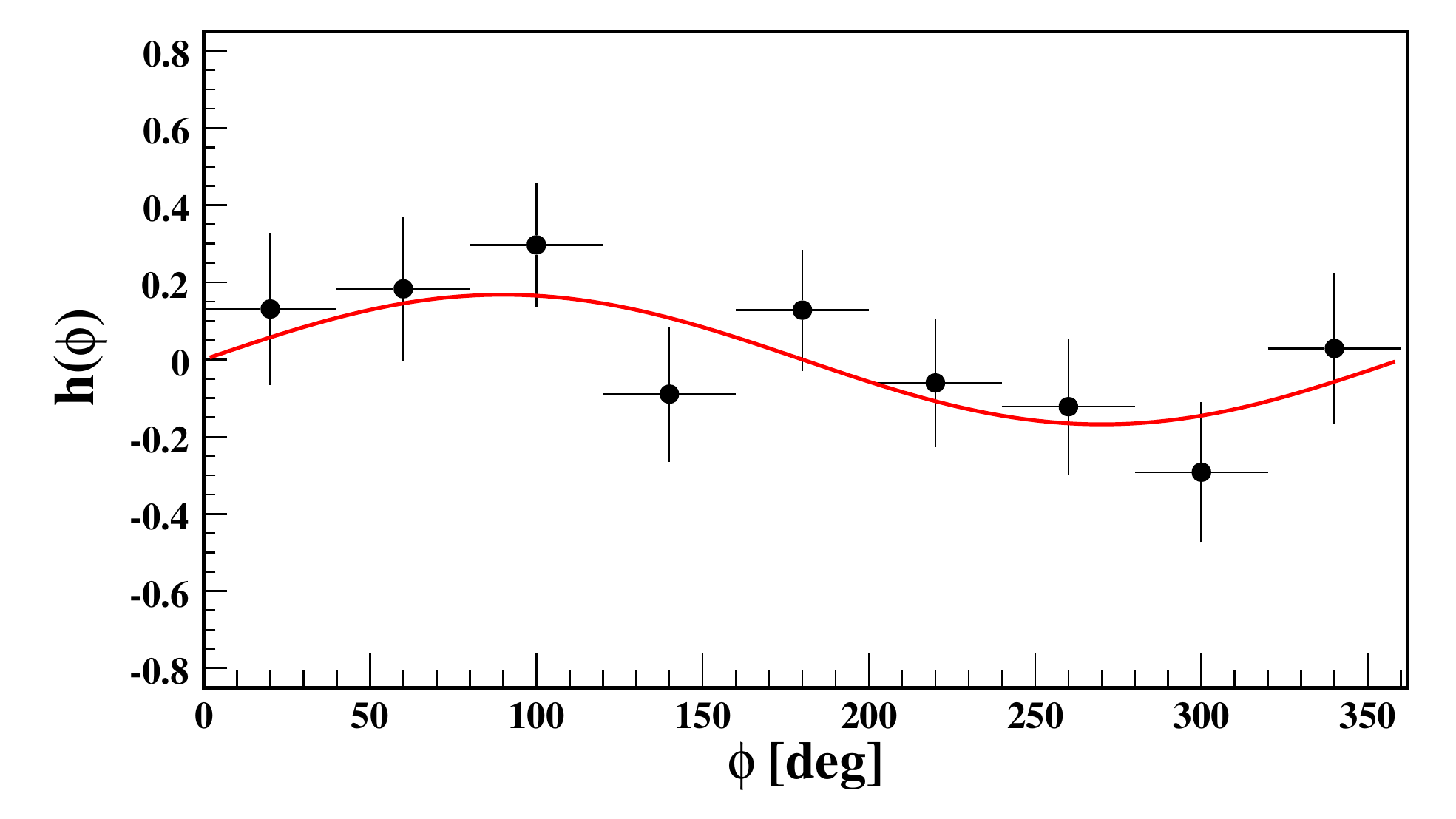} 
\captionsetup{width=.85\linewidth, justification=justified}
\caption{Asymmetry distributions of $g_{\xi ,\phi_{12}}(\phi)$ and $h_{\xi ,\phi_{12}}(\phi)$ for the configuration $\xi=(20^\circ,20^\circ, 103~{\rm MeV})$ and $\phi_{12}=100^\circ$. The error bars represent statistical uncertainties. The lines show the fit of the corresponding functions.}
\label{fig:twoo}
\end{figure*}

For a given kinematical configuration, the analyzing powers can be obtained from the following relation~\cite{Ohlsen1981}:

\begin{equation}
\begin{aligned}
N_{\xi ,\phi_{12}}^{s}(\phi) &=N_{\xi ,\phi_{12}}^{0}(\phi) (1+p_{z}^{s}A_{y}(\xi ,\phi_{12})\cos\phi \\
&-p_{z}^{s}A_{x}(\xi ,\phi_{12})\sin\phi).
\label{eq:3-4}
\end{aligned}
\end{equation}
$N^{s}$ ($N^{0}$) is the normalized number of events for a polarized (unpolarized) beam. The vector polarization of the beam is given by $p_z$ and the vector analyzing powers are indicated by $A_{x}$ and $A_{y}$. Here, $\phi$ is the angle between quantization axis for the polarization and the normal to the scattering plane of the first nucleon in the laboratory frame of reference, with $\phi_{1}=0$. Fig.~\ref{fig:coor} shows the coordinate system and the scattering plane.

$\xi$ defines a kinematical point ($\theta_{1}$, $\theta_{2}$, $S$). Since the statistics obtained with an unpolarized beam were limited, we extracted the spin observables by solely using $N^{\uparrow}$ and $N^{\downarrow}$, corresponding to the normalized number of events for the spin-up and spin-down polarized beams, respectively. The analyzing powers $A_{x}$ and $A_{y}$ are extracted using the following relation:

\begin{equation}
\begin{split}
 f_{\xi ,\phi_{12}}(\phi) & = \frac{N_{\xi ,\phi_{12}}^{{\uparrow}}(\phi)-N_{\xi ,\phi_{12}}^{{\downarrow}}(\phi)}{N_{\xi ,\phi_{12}}^{{\uparrow}}(\phi)p_{z}^{\downarrow}-N_{\xi ,\phi_{12}}^{{\downarrow}}(\phi)p_{z}^{\uparrow}} \\ 
 & =A_{y}(\xi ,\phi_{12})\cos\phi-A_{x}(\xi ,\phi_{12})\sin\phi,
\end{split}
\label{eq:4-4}
\end{equation}
where $p_{z}^{\uparrow}$ and $p_{z}^{\downarrow}$ are the values of up ($0.57\pm0.03$) and down ($-0.70\pm0.04$) beam polarization. Parity conservation imposes the following restrictions on the components of the vector analyzing powers~\cite{Ohlsen1981}:
\begin{equation}
 \begin{aligned}
 A_{x}(\xi ,-\phi_{12})= -A_{x}(\xi ,\phi_{12}); \\
 A_{y}(\xi ,-\phi_{12})= A_{y}(\xi ,\phi_{12}), 
\label{eq:three}
 \end{aligned}
\end{equation}
where for $\phi_{12}=180^{\circ}$, we expect $A_{x}=0$. By taking the sum and difference of $f_{\xi ,\phi_{12}}(\phi)$ and $f_{\xi ,-\phi_{12}}(\phi))$ in combination with the results of Eq.~\ref{eq:three}, the following combination of asymmetries for mirror configurations ($\xi$ , $\phi_{12}$) and ($\xi$ , $-\phi_{12}$) can be obtained~\cite{Stephan2010,Stephan2013}:

\begin{equation}
 \begin{aligned}
 g_{\xi ,\phi_{12}}(\phi) &= \frac{f_{\xi ,\phi_{12}}(\phi)+f_{\xi ,-\phi_{12}}(\phi)}{2},  \\
              &=A_{y}(\xi ,\phi_{12})\cos\phi.\\
 h_{\xi ,\phi_{12}}(\phi) &= \frac{f_{\xi ,\phi_{12}}(\phi)-f_{\xi ,-\phi_{12}}(\phi)}{2},  \\
              &=-A_{x}(\xi ,\phi_{12})\sin\phi.\\ 
\label{eq:8-4}
 \end{aligned}
\end{equation}

\begin{figure*}
\centering
\includegraphics[scale=0.8]{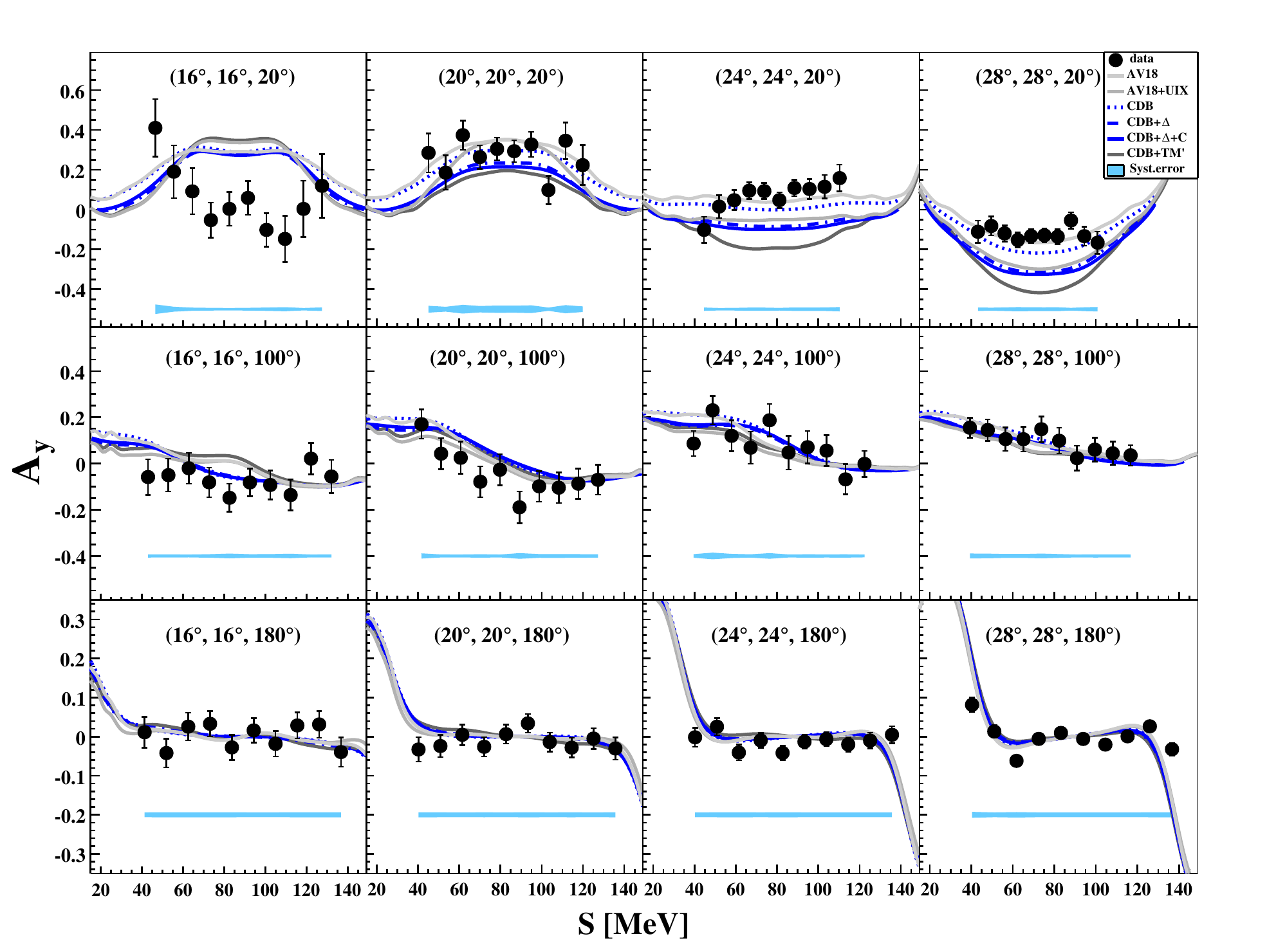}
\captionsetup{width=.85\linewidth, justification=justified}
\caption{The analyzing power, $A_{y}$, as a function of the kinematical variable $S$ for some symmetric configurations with the different $\phi_{12}$. The details of the configuration are indicated in each panel and the values refer to the angles ($\theta_1, \theta_2, \phi_{12}$). The circles are the measured $A_{y}$ and the errors are statistical only. The cyan bands show the systematic uncertainties. The lines  show the Faddeev calculations using the 2NF such as CD-Bonn (dotted line) and AV18 (light gray line) and 2NF+3NF models such as CDB+$\Delta$ (dashed-dotted line), CDB+$\Delta$ including Coulomb (solid line), CDB+TM99 (black line) and AV18+UIX (gray line)~\cite{Witala2009,witala1988,Deltuva2005,Deltuva2013,Deltuva2003,Deltuva2015}.}
\label{fig:two}
\end{figure*}

\begin{figure*}
\centering
\includegraphics[scale=0.8]{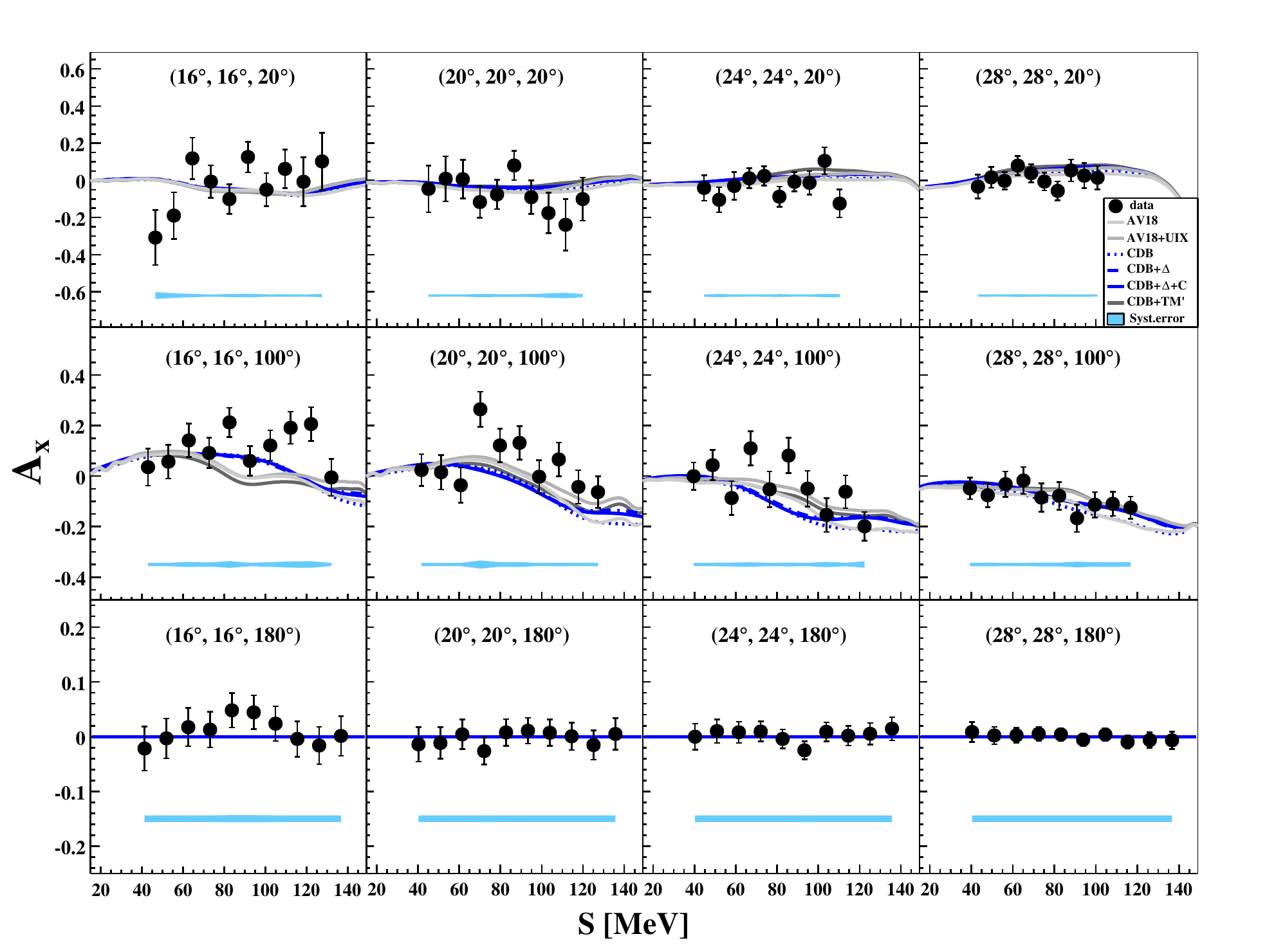} 
\captionsetup{width=.85\linewidth, justification=justified}
\caption{Same as Fig.~\ref{fig:two} except for $A_{x}$. Note that for $\phi_{12}=180^{\circ}$, the analyzing power must be exactly 0 due to parity conservation.}
\label{fig:twwoo}
\end{figure*}

\begin{figure*}
\centering
\includegraphics[scale=0.6]{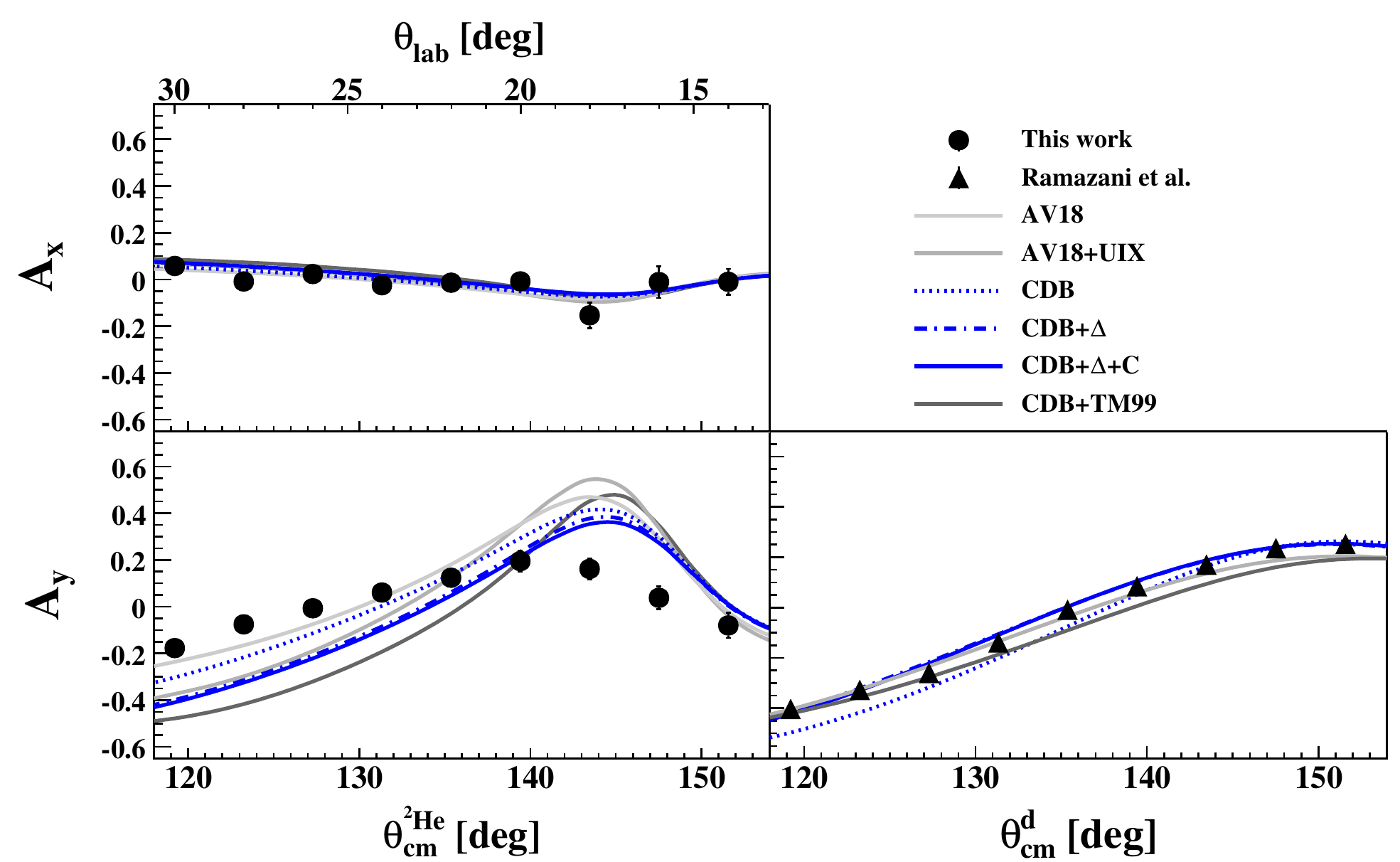}
\captionsetup{width=.85\linewidth, justification=justified}
\caption{The analyzing powers, $A_{x}$ and $A_{y}$, as a function of the center-of-mass angles for  $d(\pol{p},{\rm{^{2}He}})n$, $\theta_{cm}^{\rm{^{2}He}}$ and $d(\pol{p},d)p$, $\theta_{cm}^{d}$, at 135~MeV proton beam energy. On the left panel, the circles show the data points for $A_{x}$ and $A_{y}$ and the curves represent the results of different break-up calculation with kinematics very close to the $d(\pol{p},{\rm{^{2}He}})n$ reaction. The right panel depicts the measured $A_{y}$~\cite{Ramazani2008} and the calculations for the $\pol{p}d$ elastic reaction at 135~MeV proton beam energy.}
\label{fig:three}
\end{figure*}

By fitting the experimentally extracted polarization observables $g_{\xi, \phi_{12}}(\phi)$ and $h_{\xi, \phi_{12}}(\phi)$ with $A_{y}\cos\phi$ and $A_{x}\\sin\phi$, respectively, we extracted $A_{y}$ and $A_{x}$.
 
Fig.~\ref{fig:twoo} illustrates such a fit for the asymmetry distributions for the configuration $\xi=(20^\circ,20^\circ, 103~{\rm MeV})$ and $\phi_{12}=100^\circ$.
Figs.~\ref{fig:two} and~\ref{fig:twwoo} show our measurements of $A_{x}$ and $A_{y}$ as a function of $S$ for some symmetric configurations ($\theta_1=\theta_2$) at small, intermediate and large relative azimuthal angle $\phi_{12}$. The lines in these figures show the results of Faddeev calculations for point-like kinematical configurations in the middle range of the data using 2NFs and 2NF+3NFs as explained in the figure caption. We also performed the weighted average of the theoretical calculations in the range of the data. The results of this averaging are very close to the central values for the kinematics presented here. We have, therefore, decided to present the latter to ease the future comparisons with our data.  
The errors are statistical and the band in each panel depicts the systematic uncertainties. We identified two sources that give rise to a systematic uncertainty in the measurements of $A_y$ and $A_x$. The first one is related to the uncertainty of the beam polarization ($\sim$6\%). The second source relates to induced asymmetries due to small differences in the efficiencies and errors in the charge normalization between the data taken with the spin-up and down polarization states.  

\section{Discussion}
\label{Dis}
As observed in Figs.~\ref{fig:two} and~\ref{fig:twwoo}, for all azimuthal opening angles, the predictions based on a 2NFs are closest to the data for both analyzing powers, although, some disagreements are observed  at some places. The effects of the Coulomb force are very small for all the configurations studied in this paper. Therefore, the origin of these discrepances must lie in the treatment of 3NFs. However, when 3NF is added to the calculations, and specifically for the configurations when the azimuthal angle is small, the disagreement becomes even larger. The origin of this behaviour is not understood but was also observed in an earlier experiment with a beam energy of 190~MeV~\cite{Mardanpour2010}. 
 
The analyzing powers of the reaction $d(\pol{p},pp)n$ were measured for symmetic configurations of two protons,  $\theta_{1}=\theta_{2}$, with small opening angle $\phi_{12}=20^{\circ}$ and small relative energy of less than 1~MeV at the center of $S$-curve. These configurations are similar to the configuration of $d(\pol{p},{\rm{^{2}He}})n$. The analyzing powers for the corresponding reaction, $d(\pol{p},{\rm{^{2}He}})n$, can be compared to the analyzing powers of the elastic $d(\pol{p},d)p$ scattering. In the elastic channel, the total isospin of the initial and final state is exclusively $\frac{1}{2}$, whereas in the former case, the final state could couple to an isospin $\frac{3}{2}$ as a consequence of the isospin violating Coulomb force~\cite{Mardanpour2010}. To obtain the values of $A_{x}$ and $A_{y}$ corresponding to $d(\pol{p},{\rm{^{2}He}})n$ we use a second-order polynomial fit to the measured analyzing powers as a function of $S$ for $\phi_{12}=20^\circ$. We took the fit value and its error at the central value of $S$, corresponding to the smallest relative energy. 
In these cases, since the two protons move very close to each other, the most probable wave function would have an angular momentum $L=0$ and isospin $I=1$~\cite{Mardanpour2010} lending confidence to the fact that this state carries the quantum numbers of $\rm{^{2}He}$.  
We extracted the corresponding predicted values for $A_{x}$ and $A_{y}$ from theory using the central value of $S$, with the smallest relative energy of two protons. Fig.~\ref{fig:three} shows the values of $A_{x}$ and $A_{y}$ as a function of center-of-mass angles corresponding to $d(\pol{p},{\rm{^{2}He}})n$, $\theta_{cm}^{\rm{^{2}He}}$ and $d(\pol{p},d)p$, $\theta_{cm}^{d}$. In the top left panel of Fig.~\ref{fig:three}, one can observe that the measured values of $A_{x}$ are very close to zero as required for a two-body reaction.  The filled circles in the bottom left panel of Fig.~\ref{fig:three} show the data for $A_{y}$.  For center-of-mass angles less than $135^{\circ}$ the experimental data points are higher than the values of all 2N+3NF theoretical calculations while for center-of-mass angles larger than $135^{\circ}$, our data fall below that of the calculations, however the overall tendency seems to be well reproduced. Fig.~\ref{fig:four} compares our results for $A_{y}$ for $d(\pol{p},{\rm{^{2}He}})n$ with those taken at the higher beam energy. The circles and squares depict the data points of $A_{y}$ at 135~MeV and 190~MeV, respectively. The solid and dashed lines show the CDB+$\Delta$+Coulomb predictions at 135~MeV and 190~MeV proton beam energies, respectively. For both proton beam energies there is a disagreement between the data and the calculations including a 3NF.

\section{Summary and conclusion}
\label{summary}
The need for an additional 3NF became clear through studying three-nucleon scattering processes.
To investigate systematically 3NF effects in three-nucleon scattering processes, the $\pol{p}+d$ break-up reaction was studied using a polarized beam of protons impinging on a liquid deuterium target. Our main priority was to study configurations in which both protons scatter to polar angles smaller than $30^{\circ}$ and with a small, intermediate and large relative azimuthal opening angle. Analyzing powers have successfully been measured and are presented as a function of $S$ for different combinations of ($\theta_{1}=\theta_{2}$, $\phi_{12}$). The major discrepancies between the data and the theoretical calculations arise at small azimuthal opening angles, $\phi_{12}=20^{\circ}$, which corresponds to $d(\pol{p},{\rm{^{2}He}})n$. In this range, the predictions based a 2NF are close to the data, although, the disagreement is still significant. The inclusion of 3NFs increases the gap between data and predictions.  

\begin{figure}
\centering
\includegraphics[scale=0.43]{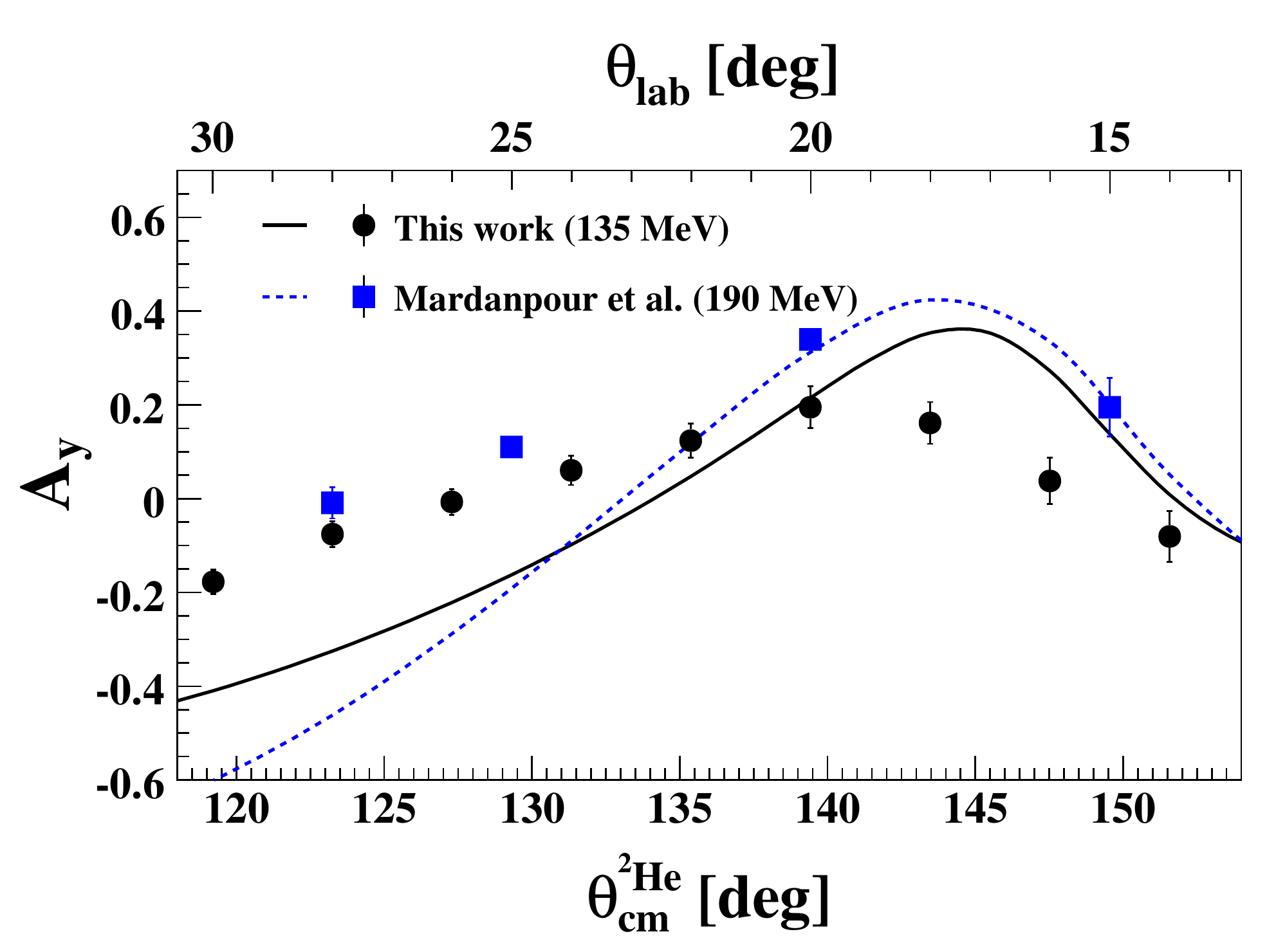} 
\caption{$A_{y}$ versus the center-of-mass angles for $d(\pol{p},{\rm{^{2}He}})n$ reaction at two proton beam energies. The circles and squares depict the data points of $A_{y}$ for proton beam at 135~MeV and 190~MeV, respectively. The solid and dashed lines show the CDB+$\Delta$+Coulomb predictions at 135~MeV and 190~MeV proton beam energies, respectively.}
\label{fig:four}
\end{figure} 

The results of $d(\pol{p},{\rm{^{2}He}})n$ for 135~MeV proton beam energy are compared to the results using a proton beam with an energy of 190~MeV~\cite{Mardanpour2010}.  A comparison between two proton beam energies of 135~MeV and 190~MeV in Fig.~\ref{fig:four} shows that the spin-isospin deficiency that was observed before does not diminish at lower energies.

%
\section*{Acknowledgement}
This work was partly supported by the Polish National Science Centre under Grant No. 2012/05/B/ST2/02556 and 2016/22/M/ST2/00173. The numerical calculations of were partially performed on the supercomputer cluster of the JSC, J\"ulich, Germany.
%
\bibliographystyle{epj}
\bibliography{MyPaper}

\end{document}